\documentclass[twocolumn]{aastex63}

\newcommand{\HST}{{\it HST}}

\received{May 20, 2025}
\revised{June 22, 2025}
\accepted{June 24, 2025} 
\submitjournal{ApJL}

\usepackage{xcolor}
\usepackage{enumerate}

\let\a=\alpha \let\b=\beta \let\g=\gamma \let\d=\delta \let\e=\epsilon
\let\h=\eta  \let\k=\kappa \let\l=\lambda 
\let\s=\sigma \let\t=\tau \let\u=\upsilon
\let\r=\rho
 
\let\w=\omega	   \let\G=\Gamma \let\D=\Delta  \let\L=\Lambda
\let\S=\Sigma \let\W=\Omega
\let \x = \chi \let \z = \xi

\def\be{\begin{eqnarray}}
\def\ee{\end{eqnarray}}
\def\ba{\begin{array}}
\def\ea{\end{array}}

\def \Msun {M_{\odot}}

\def \pc {\mathrm{pc}}

\def \kms {\mathrm{km \, s^{-1}}}
\def \cm {\mathrm{cm}}
\def \M  {\mathscr{M}}
\def \K {\mathrm{K}}
\def \sec {\mathrm{s}}
\def \gramm {\mathrm{g}}

\def \earth {\oplus}
\def \S {Section }

\begin{document}

\title{Observability of Isolated Stellar-mass Black Holes}

\correspondingauthor{Lena Murchikova, Kailash Sahu}
\email{lena@northwestern.edu, ksahu@stsci.edu}

\author[0000-0001-8986-5403]{Lena Murchikova}
\affiliation{Center for Interdisciplinary Exploration and Research in Astrophysics, Northwestern University, Evanston, IL 60208, USA}
\affiliation{Department of Physics \& Astronomy, Northwestern University, Evanston, IL 60208, USA}
\affiliation{School of Natural Sciences, Institute for Advanced Study, 1 Einstein Drive, Princeton, NJ 08540, USA}

\author[0000-0001-6008-1955]{Kailash C. Sahu}
\affiliation{Space Telescope Science Institute, 3700 San Martin Drive, Baltimore, MD 21218, USA}
\affiliation{Eureka Scientific Inc., 2542 Delmar Avenue, Suite 100, Oakland, CA 94602-3017, USA}


\begin{abstract}
Stellar-mass black holes (BHs) represent the natural end states of massive stars. It is estimated that $10^8$ stellar-mass BHs are present in the Milky Way galaxy, a significant fraction of which are expected to be isolated. Despite their expected abundance, only about 20 have been detected so far---mostly in binary systems---with just one confirmed isolated black hole (IsoBH) identified via astrometric microlensing. In this study, we investigate the potential for detecting electromagnetic emissions from IsoBHs by generating synthetic model spectra of their emissions in different types of interstellar medium environments.
These model spectra are then compared with current observational capabilities.
We show that photons emitted by IsoBHs---
especially those accreting material in dense environments or within the Solar neighborhood—--should be readily detectable. However, confidently identifying these sources remains highly challenging. We conclude that a number of IsoBHs must already exist in current astronomical catalogs but have not been identified as such. We outline possible strategies for detection and identification of IsoBHs using the current and upcoming telescopes.
\end{abstract}



\section{Introduction} \label{sec:intro}

Stellar-mass black holes (BHs) represent the natural end state of massive stars \citep{1939PhRv...55..364T,1939PhRv...55..374O,2015PASA...32...16S}. It is estimated that about $10^8$ stellar-mass BHs are present in the Milky Way galaxy \citep{1992Vdheuvel,Brown1994,Samland1998}, a substantial fraction of which are expected to be single, rather than belonging to binaries [see \citet{Sahu2022}].
Observationally we have found only about twenty BHs in binaries with stellar companions \citep{2016ApJS..222...15T,2016A&A...587A..61C, elbadry2024}, and only one (OGLE-2011-BLG-0462/MOA-2011-BLG-191) isolated black hole (IsoBH), found using an extensive microlensing survey spanning over a decade and multiple epochs of \HST\ observations \citep{Sahu2022, Sahu2025, Lam2023}. \citet{Kimura2025} explored the detectability of electromagnetic radiation from this IsoBH.

General first-principle discussions of observability of IsoBHs dates back several decades \citep[e.g.,][]{1979grae.book.....H}. 
In this work we explore current observability of electromagnetic emission from IsoBHs by generating synthetic model spectra of a typical IsoBH moving through various types of interstellar medium (ISM): warm ionized and neutral medium, cold neutral medium, molecular clouds, and hot ionized medium (or coronal gas). We compare the model spectra with current observational capabilities of the Square Kilometer Array (SKA), the Atacama Large Millimiter/Submillimiter Array (ALMA), James Webb Space Telescope (JWST), and Chandra Space Telescope. For the purposes of this paper any stellar mass BH which is not in a close binary system is considered an IsoBH.

The only IsoBH detected so far, OGLE-2011-BLG-0462, has a mass of $7.15\pm0.83\, M_\odot$, and a spacial velocity  of $51.1\pm7.5 \kms$ in the plane of the sky relative to the stars in its neighborhood \cite{Sahu2025}. 
Masses of nearly two dozen BHs detected in X-ray binary systems in our  Galaxy show a distribution peaking near $7–8\, M_\odot$, and ranging
from $\sim 5$ to 20~$M_\odot$ \citep{ozel2010, 2016A&A...587A..61C,miller-jones2021}. 
Recently, Gaia astrometric data have been used to detect three BHs orbiting normal stars in non-interacting binaries with  masses of 9.3, 9.0 and 33 $M_\odot$ 
\citep{2023MNRAS.518.1057E, 2023AJ....166....6C, 2023MNRAS.521.4323E, Gaia2024}.
The space velocities of the BHs are uncertain but 
generally range from $\sim 10$ to $\sim 200 \, \kms$ \citep{mandel2016,repetto2012}. Observations of Gravitational waves by the Laser Interferometry Gravitational Wave Observatory (LIGO)
and Virgo Collaboration  \citep{Abbott2023} have led to the detection of nearly 100 merging BHs in external galaxies, with masses ranging from $\sim$ 5 to 150  $M_\odot$. 

In our analysis here, we primarily assume a prototypical IsoBH with a mass $M_\bullet$ and velocity $v_\bullet$ with respect to the surrounding medium:
\begin{eqnarray}
M_\bullet = 10 \, M_\sun, \quad 
v_{\bullet} = 50 \, \kms \label{eq:mv}.
\end{eqnarray}
Both parameters are similar to those reported by \cite{Sahu2025} for the only observed IsoBH, OGLE-2011-BLG-0462. They are also close to the expected peak of the mass distribution of the galactic BHs — though somewhat more rounded — and fall within the mid-range of expected velocities. We discuss the uncertainties arising from variations in mass and velocity later in the paper.

The paper is organized as follows. In Section 2, we briefly discuss simple BH accretion theory and the model used to generate the spectra. In Section 3, we estimate Bondi accretion rates for an IsoBH moving through all the most common types of ISM. In Section 4, we focus on the types of ISM present at different distances from the Sun and 
assess the observability of IsoBHs accreting from these environments using current telescopes.
In Section 5, we discuss the strategies and challenges in identifying IsoBHs based on their electromagnetic emissions. 
In Section 6, we summarize our conclusions and discuss future prospects.

\section{Accretion onto Isolated BH}\label{sec:accr}

The Bondi mass accretion model \citep{1952MNRAS.112..195B,1944MNRAS.104..273B} assumes uniformity of the accreting medium, spherical symmetry of the accretion flow, no BH feedback, and absence of rotation which would slow the accretion down. So it can only serve as an upper limit on the accretion rate onto a BH. 
Yet, this upper limit is very useful in estimating the type of the accretion flow onto the BH.

The Bondi mass accretion onto an IsoBH moving through ISM can be written as \citep{1985MNRAS.217..367S}
\be\label{eq:bondi}
\dot{M}_\mathrm{B}&=& 4\pi\frac{(GM_\bullet)^2 }{ ( v_\bullet^2+ C_\mathrm{ISM}^2)^{3/2}} \mu_\mathrm{ISM} n_\mathrm{ISM} m_p,
\ee
where $G$ is a gravitational constant, $M_\bullet$ is the mass of the BH, $\mu_\mathrm{ISM}$ is the mean atomic mass of the ISM (so that $\mu_\mathrm{ISM} \simeq 1$ for neutral medium and $\mu_\mathrm{ISM}\simeq 0.5$ for ionized medium), $n_\mathrm{ISM}$ is the number density of the ISM, $m_p$ is the proton mass, $C_\mathrm{ISM}$ is the sound speed in the ISM through which the BH is moving, and $v_\bullet$ is velocity of the BH relative to the ISM.

For convenience we introduce the parameter $\lambda$ to denote the near-horizon accretion rate of the IsoBH in units of Bondi accretion
\be
    \dot{M}_\bullet = \lambda  \dot{M}_\mathrm{B}, \quad \lambda \leq 1. \label{eq:lambda}
\ee
The parameter $\lambda$ must not exceed 1, since the IsoBH cannot accrete more than it gravitationally captures. 

We can estimate the speed of sound in the ISM from first principles assuming that the ISM is an ideal gas 
\be \label{eq:sound}
C_\mathrm{ISM} \simeq \sqrt{\gamma \frac{p_\mathrm{ISM}}{\rho_\mathrm{ISM}}} =  \sqrt{\gamma \frac{k T_\mathrm{ISM}}{\mu_\mathrm{ISM} m_p}},
\ee 
where $\gamma$ is the adiabatic constant,  $\rho_\mathrm{ISM}$ is the density, $p_\mathrm{ISM}$ is the pressure, and $T_\mathrm{ISM}$ is the temperature of the ISM.

We see that for all the states of the ISM with $T_\mathrm{ISM} \leq 10^4 \, \K$, i.e. all except the hot ionized (coronal) gas (Section \ref{sec:3}), the speed of sound in the ISM is considerably smaller than the IsoBH velocity. So the denominator in Equation \ref{eq:bondi} for all except the case of coronal gas can be simplified as
\be\label{eq:simp_not-hot}
    ( v_\bullet^2+ C_\mathrm{ISM}^2)^{3/2} \simeq v_\bullet^3.
\ee
With substitution \ref{eq:simp_not-hot}, we find that 
\be
\dot{M}_\mathrm{B}= 0.89 &\times& 10^{11} M_{\bullet,10}^2 v_{\bullet,50}^{-3} \nonumber \\
&\times&
\left[\frac{\mu_\mathrm{ISM}}{1}\right]\left[\frac{n_{{\mathrm{ISM}}} }{0.3 \, \cm^{-3}} \right]  \, \gramm \, \sec^{-1}, \label{eq:Bnum}
\ee
where $M_{\bullet,10}=\frac{M_{\rm \bullet}}{10\,M_\odot},$ $v_{\bullet,50}=\frac{v_\bullet}{50 \, \kms},$
and for scaling purposes we used a typical density of the warm ionized/neutral medium $0.3 \ \cm^{-3}$ \citep{2001RvMP...73.1031F}.

Let us now compare this value with the Eddington accretion rate
\be
\dot{M}_\mathrm{Edd}=\frac{4\pi G M_\bullet m_p}{\eta c \sigma_T} = 1.40 \times 10^{19} M_{\bullet,10}  \ \gramm \, \sec^{-1},
\ee
where $c$ is the speed of light, $\sigma_T$ is the Thomson cross-section, 
and $\eta$ denotes the radiation efficiency, defined as the fraction of accreted $mc^2$ radiated by the BH. In the calculations that follow, we adopt $\eta=0.1$.
Note that $\eta$ is distinct from $\lambda$ defined in Equation \ref{eq:lambda}. 
The later is simply the fraction of Bondi accretion rate of the BH that is reaching the horizon ($\dot{M}_\bullet=\lambda \dot{M}_\mathrm{B}$). 
The former implies that if the BH's accretion rate at the horizon is $\dot{M}_\bullet,$ then over a time interval $\delta t$, the BH  will accrete a mass $\dot{M}_\bullet \delta t$ and radiate an energy $\eta \dot{M}_\bullet c^2 \delta t.$

We find that the near-horizon accretion rate of the IsoBH $(\dot{M}_\bullet)$ is many orders of magnitude smaller than the Eddington accretion rate
\be
    \frac{\dot{M}_\bullet}{\dot{M}_\mathrm{Edd}} \leq \frac{\dot{M}_\mathrm{B}}{\dot{M}_\mathrm{Edd}}
    = 6.35&\times&10^{-9} M_{\bullet,10} v_{\bullet,50}^{-3} \nonumber
    \\
    &\times&
    \left[\frac{\mu_\mathrm{ISM}}{1}\right]
    \left[\frac{n_{{\mathrm{ISM}}} }{0.3 \, \cm^{-3}} \right].
\ee
This remains true even for an IsoBH accreting ISM from the core of a giant molecular clouds (GMC), which is the densest medium we consider. So we can conclude that every IsoBH is underfed and consequently accretes via radioactively inefficient accretion flow, also referred to as hot accretion flow \citep{1977ApJ...214..840I,1994ApJ...428L..13N,2014ARA&A..52..529Y}.

The most studied BH with radioactively inefficient accretion flow is the Milky Way's Galactic center BH Sagittarius A* \citep{2018A&A...615L..15G,2019Sci...365..664D}. When calculating the spectrum of our typical IsoBH below, we will be guided by this example, with due consideration to the difference in mass. One can show that the relative environmental conditions around these two underfed black holes are similar (e.g. \citealt{Kimura2025}). We assume that the near-horizon accretion rate of the IsoBH, similar to that of Sagittarius A*, is approximately 1\% of the Bondi accretion rate \citep{Genzel2010}:
\be
    \lambda = 0.01.
\ee
For convenience of the analysis, we also include in our plots the estimated IsoBH spectrum for $\lambda = 1$, corresponding to 100\% Bondi rate, which can be thought of as an upper limit.
In order to compute the spectrum of the IsoBH, we use the Low-Luminosity AGN Spectral Energy Distribution code \texttt{LLAGNSED}, which is based on the analytical framework 
proposed by \cite{1997ApJ...477..585M} and subsequently updated and publicly released by \cite{2021ApJ...923..260P}.
We use the standard parameters listed by \citet{2021ApJ...923..260P}, such as viscosity $\alpha = 0.2,$ and plasma $\beta =10.$
The code was written for applications to the supermassive BH, but can be applied to any case of an underfed BH with a hot accretion flow around it.

\begin{figure*}
\vspace{-0.0cm}
\centering
\begin{tabular}{cc}
\includegraphics[width=0.99\columnwidth]{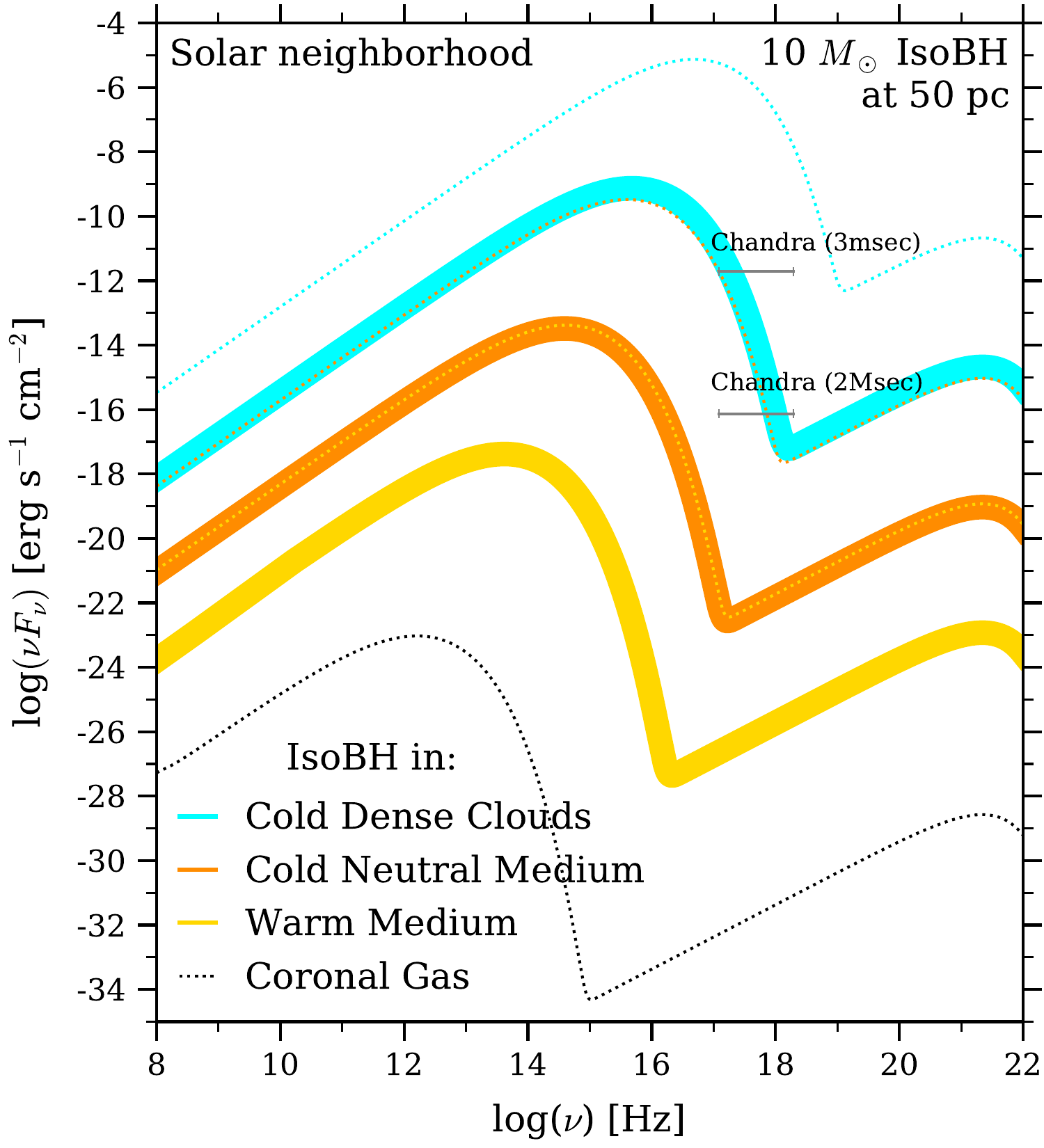} & 
\includegraphics[width=0.99\columnwidth]{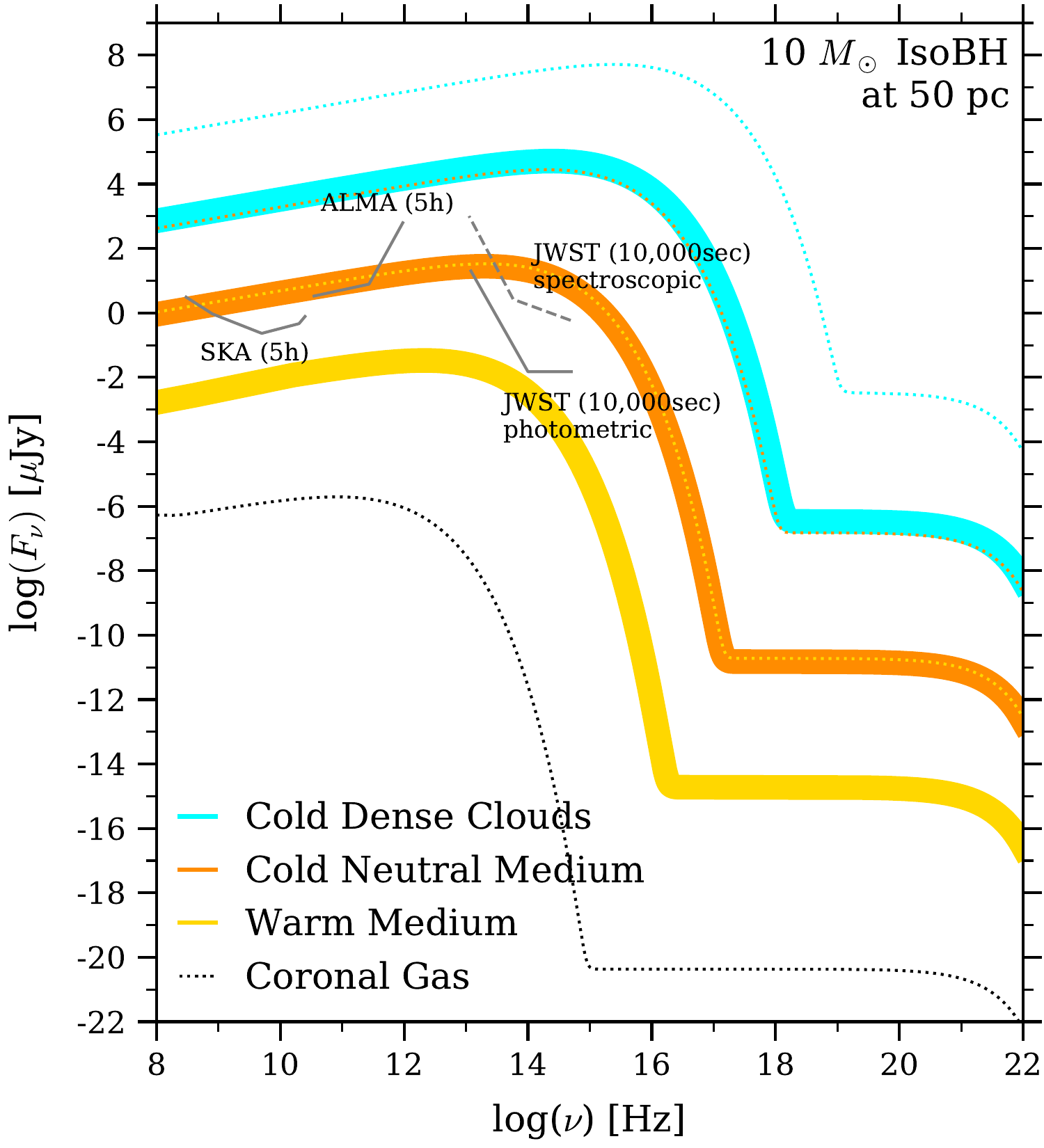}
\end{tabular}
\caption {Estimated emission spectra of an IsoBH located 50 pc from the Sun, accreting gas from four types of local ISM: cold dense cloud with a density of $\sim 3,000 \, \cm^{-3}$ (cyan), clumps within local interstellar clouds with a density of $\sim 20 \, \cm^{-3}$ (orange), warm partially ionized medium with a density of $\sim 0.3 \, \cm^{-3}$ (yellow), and hot ionized coronal gas (grey). Other types of Galactic ISM are not present within 50 pc from the Sun.
The thin dotted lines show the expected emission spectrum from an IsoBH accreting at the Bondi rate $(\lambda=1)$, representing the upper limit of the emission. The thick solid lines below the dotted lines, shown in corresponding colors, represent the spectrum from the IsoBH accreting at a more realistic 1\% of the Bondi rate $(\lambda=0.01)$. The exaggerated thickness illustrates large uncertainties in the accretion-model parameters. 
Note that some of the thin dotted lines overlap with the thick solid lines of other colors.
We only plot the upper limit for the coronal gas corresponding to the Bondi rate, as the emission is already too low to be detectable. 
For convenience, we show the spectra in two different units ($\nu F_\nu$ and $F_\nu$) commonly used in different wavelength regions. Dark gray lines indicate the observational sensitivities of Chandra, JWST, ALMA, and SKA, achievable within the integration times listed in brackets. 
} 
\label{fig:solar50}
\end{figure*}

\section{Isolated BH in ISM}\label{sec:3}

In this Section we estimate Bondi accretion rates for an IsoBH moving through the most common types of ISM 
\citep{1977ApJ...218..148M,2011piim.book.....D}.

\subsection{Warm Ionized/Neutral Medium}

The warm ionized medium and the warm neutral medium (WM) are the most abundant states of the ISM in the Galaxy \citep{2001RvMP...73.1031F} by volume. Typical densities of such ISM are $\sim 0.2-0.5 \, \cm^{-3},$ and temperatures can be between $6,000\, \K$ and $10,000 \, \K.$ 

From Equation \ref{eq:Bnum} we find that an IsoBH accreting from warm ionized/neutral medium has a Bondi accretion rate of 
\be\nonumber
\dot{M}_\mathrm{B}\big|_\mathrm{WM} \simeq  6.66 &\times& 10^{10} M_{\bullet,10}^2 v_{\bullet,50}^{-3}\\
&\times&
\left[\frac{\mu_\mathrm{WM}}{0.75}\right]\left[\frac{n_{{\mathrm{ISM}}} }{0.3 \, \cm^{-3}} \right]  \, \gramm \, \sec^{-1},
\label{eq:BMW}
\ee
where we set $\mu_{WM} = \frac{1}{2}(0.5+1) \simeq 0.75,$ i.e. the average between ionized and neutral cases.

\subsection{Cold Neutral Medium}

Cold Neutral Medium, also referred to as cold clumps, constitute a few percent of the ISM in the Galaxy by volume. It is composed of neutral gas with densities $\sim 20 - 50 \, \cm^{-3},$ and temperatures of $\sim 100 \, \K$ \citep{2001RvMP...73.1031F}. 

From Equation \ref{eq:Bnum} we find that an IsoBH accreting from cold neutral medium has a Bondi accretion rate of 
\be\nonumber
\dot{M}_\mathrm{B}\big|_\mathrm{CNM}\simeq 5.92 &\times& 10^{12} M_{\bullet,10}^2 v_{\bullet,50}^{-3}\\
&\times&
\left[\frac{\mu_\mathrm{CNM}}{1}\right]\left[\frac{n_{{\mathrm{ISM}}} }{20 \, \cm^{-3}} \right]  \, \gramm \, \sec^{-1}.
\label{eq:BCNM}
\ee

\subsection{Molecular Clouds}

Molecular Clouds (MCs) comprise only about 1\% of the ISM by volume. 
They are characterized by dense cores surrounded by more diffuse outer regions. Core densities typically reach around  $\sim 10^4 \, \cm^{-3}$, but in most extreme cases---such as in Giant Molecular Clouds (GMCs)---densities can reach up to $\sim 10^6 \, \cm^{-3}.$ However, the volume occupied by the cores is small---only 0.1 to 2\% of the total volume of the MC. The typical average density of a MC is $\sim 10^3 \, \cm^{-3}$ \citep{2020SSRv..216...76B,Pattle2017}.

The Bondi accretion rate onto the IsoBH moving through a typical MC, and the core of a GMC (GMCC) can be written as 
\be\nonumber
\dot{M}_\mathrm{B}\big|_\mathrm{MC}\simeq 2.96 &\times& 10^{14} M_{\bullet,10}^2 v_{\bullet,50}^{-3}\\
&\times&
\left[\frac{\mu_\mathrm{ISM}}{1}\right]\left[\frac{n_{{\mathrm{ISM}}} }{10^3 \, \cm^{-3}} \right]  \, \gramm \, \sec^{-1},
\label{eq:MC}
\\
\nonumber
\dot{M}_\mathrm{B}\big|_\mathrm{GMCC}\simeq 2.96 &\times& 10^{17} M_{\bullet,10}^2 v_{\bullet,50}^{-3}\\
&\times&
\left[\frac{\mu_\mathrm{ISM}}{1}\right]\left[\frac{n_{{\mathrm{ISM}}} }{10^{6} \, \cm^{-3}} \right]  \, \gramm \, \sec^{-1}.
\label{eq:GMCC}
\ee
Bondi accretion rate from the cores of ordinary MCs and the
peripheral regions of GMCs can be derived 
from the above expressions through simple rescaling.

\subsection{Coronal Gas}

Coronal Gas (CG), also known as hot ionized medium, is one of the most abundant states of the ISM in the Galaxy \citep{1977ApJ...218..148M}. It has a low density ($n_\mathrm{CG} \sim 10^{-4} - 10^{-2} \, \cm^{-3}$) and high temperature ($T_\mathrm{CG} = 10^6 - 10^7 \, \K).$ Such hot gas is present both in Galactic corona and inside the Galaxy as it can be produced by shock heating in supernova explosions. It's a special case for our calculations, since
the speed of sound in this case dominates the denominator in Equation \ref{eq:bondi}. Using Equation \ref{eq:sound} we can estimate the speed of sound in the coronal gas as
\be
    C_\mathrm{CG} \sim 240 \, \kms,
\ee
which is much larger than our assumed $v_\bullet$ and so the IsoBH's accretion rate for such an IsoBH would reduce to
\be\nonumber
    \dot{M}_\mathrm{B}\big|_\mathrm{CG}\simeq 2.31 &\times& 10^{6} M_{\bullet,10}^2 
\left( \frac{C_\mathrm{CG}}{200\kms} \right)^{-3}\\
&\times&
\left[\frac{\mu_\mathrm{ISM}}{1/2}\right]\left[\frac{n_{{\mathrm{ISM}}} }{10^{-3} \, \cm^{-3}} \right]  \, \gramm \, \sec^{-1},
\ee
i.e. spherical Bondi accretion. Recent works \citep{2022MNRAS.512.2154G,2024ApJ...966..103P} show that this remains true for any subsonic accretion with $v_\bullet < C_\mathrm{CG}$.

\begin{figure*}
\vspace{-0.0cm}
\centering
\begin{tabular}{cc}
\includegraphics[width=0.99\columnwidth]{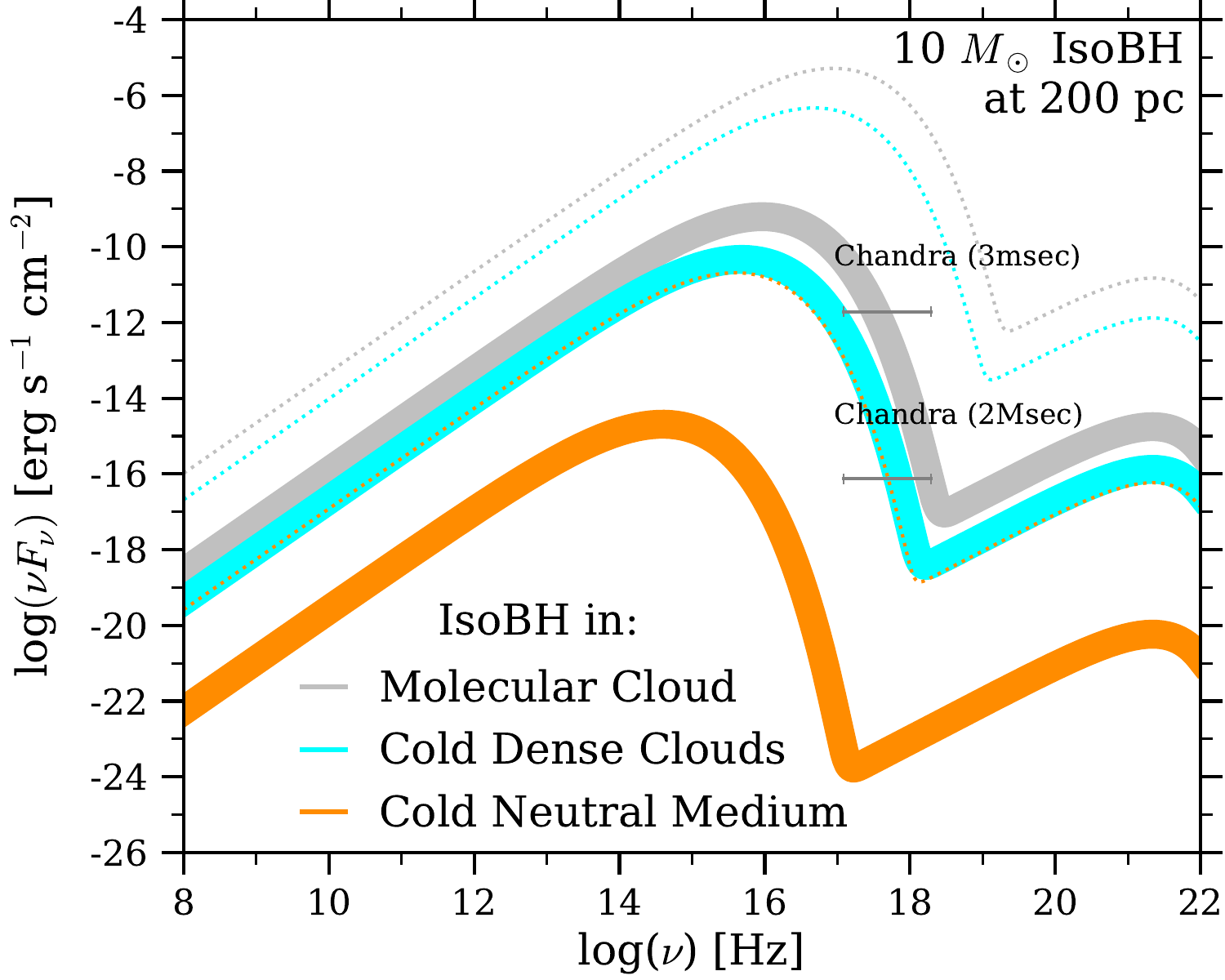} & 
\includegraphics[width=0.99\columnwidth]{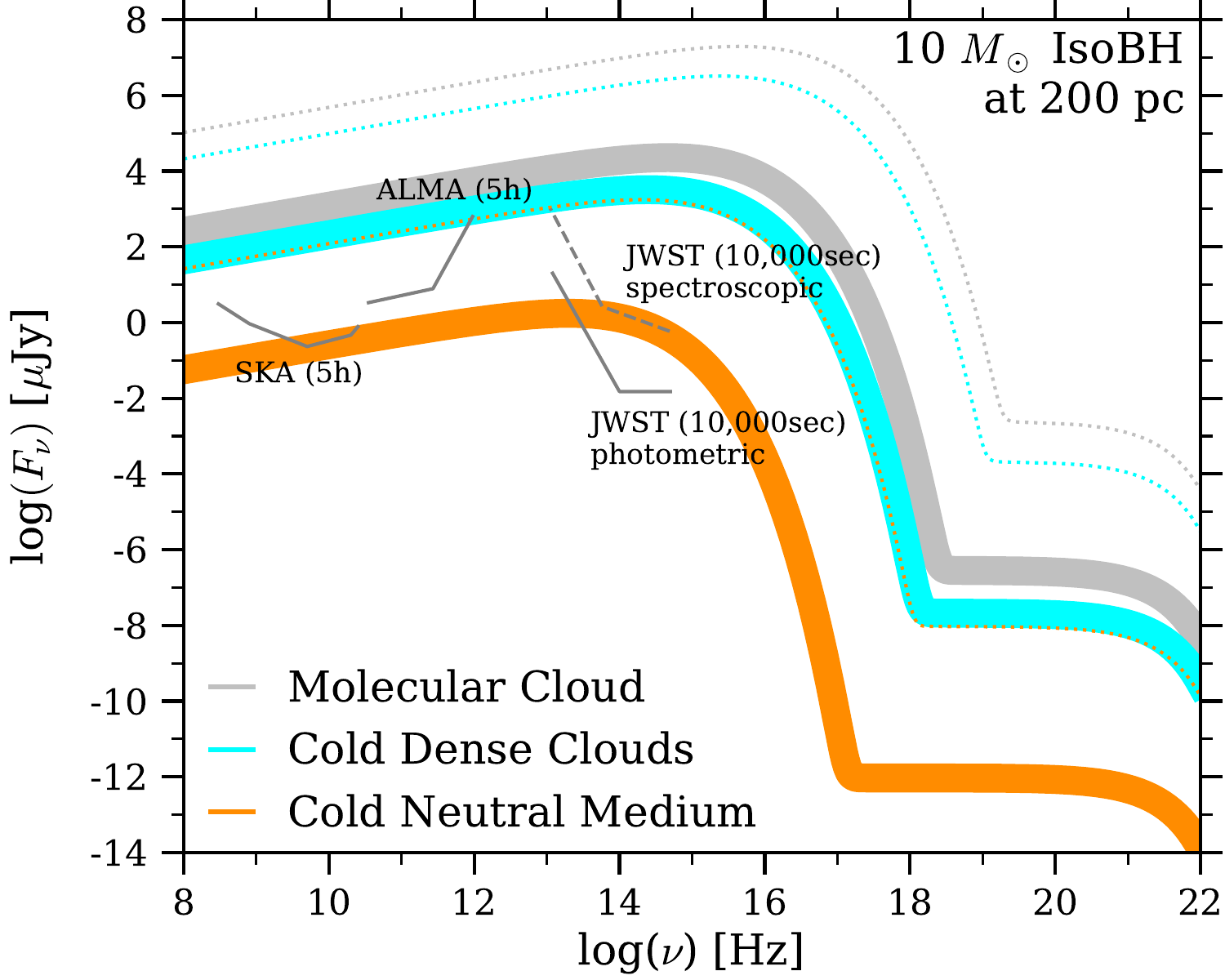}
\end{tabular}
\caption {
Estimated emission spectra of an IsoBH located 200 pc from the Sun, accreting gas from three types of local ISM: 
MCs with an average density within the core of $\sim 10^4 \, \cm^{-3}$ (grey), cold dense clouds with a density of $\sim 3,000 \, \cm^{-3}$ (cyan), and
gas clumps within local interstellar clouds with a density of $\sim 20 \, \cm^{-3}$ (orange). Other types of medium, such as warm medium and coronal gas, are also present within 200 pc from the Sun. However, a $10 \Msun$ IsoBH located at 200 pc in such a medium would be undetectable  (as evident from Figure \ref{fig:solar50}), hence we do not show them here.
The thin dotted line shows the expected emission spectrum from an IsoBH accreting at the Bondi rate $(\lambda=1)$, representing the upper limit of the emission. The thick solid lines below the dotted
lines, shown in corresponding colors, represent spectra from the IsoBH accreting at a more realistic 1\% of the Bondi rate $(\lambda=0.01)$. The exaggerated thickness illustrates large uncertainties in the accretion-model parameters. Note that some of the thin dotted orange line overlaps with the thick solid cyan.
For convenience, we show the spectra in two different units ($\nu F_\nu$ and $F_\nu$) commonly used in different wavelength regions. Dark gray lines indicate the observational sensitivities of Chandra, JWST, ALMA, and SKA, achievable within the integration times listed in brackets.}
\label{fig:solar200}
\end{figure*}

\section{Detectability} \label{sec:detect}

\subsection{IsoBH at 50 pc from the Sun}

The total stellar mass of the Milky Way galaxy is $6.08 \pm 1.14 \times 10^{10} \Msun$ \citep{Licquia2015}. 
As described in Section 1, the total number of IsoBHs in the Galaxy is estimated to be $10^8$.
Hence we can expect one IsoBH per $\sim 600 \, \Msun$ of stellar mass.
The stellar mass density in the solar neighborhood is $\sim 0.04 \Msun \pc^{-3}$ \citep{2017MNRAS.470.1360B}, which implies that there is on the average one IsoBH within each $15,000 \, \pc^3$. This corresponds to one IsoBH within a sphere of radius $\sim 15$ pc.

We can therefore expect about 35 IsoBHs within 50 pc of the Sun. In this region, there are two dominant classes of ISM: the warm partially ionized medium of the Local Interstellar Cloud, and the hot ionized gas known as coronal gas. Within warm medium, we also have clumps of cold neutral medium with densities of $\simeq 20 \, \cm^{-3}$ and temperatures of $\sim 8000 \, \K$ \citep{2011ARA&A..49..237F}. In addition, there are cold dense clouds---the closest and largest of which is the Local Leo Cold Cloud, located at a distance of $11-45 \, \pc$, with a density $n_\mathrm{CC} = 3,000 \, \cm^{-3}$ and a temperature $T_\mathrm{CC} \simeq 20 \, \K$ \citep{2012ApJ...752..119M}. The group of these local clouds is called the Local Ribbon of Cold Clouds \citep{2010A&A...514A..27H}. Such clouds are expected to be present in the ISM across the Galaxy.

Figure \ref{fig:solar50} shows the estimated emission from a $10 \Msun$ IsoBH moving through the ISM with $50 \, \kms$ 
located at a distance of $50 \, \pc$ from the Sun and accreting from four main types of ISM present within 50 pc from the Sun: cold dense clumps (cyan), cold neutral medium (orange), warm medium (yellow), and coronal gas (grey).

Comparing the estimated spectra with the observational sensitivities of Chandra, SKA, ALMA, and JWST, it becomes clear that detecting emission from an IsoBH located in the coronal gas—even relatively close to the Sun—is highly unlikely. Detection would only be feasible if the IsoBH were within 1 parsec of the Sun. However, if it were that close, it would likely have already been identified through other means, such as its gravitational influence. 

The emission from an IsoBH located in the warm medium of the local interstellar cloud can be detected with several hours of integration using SKA, ALMA, or JWST. 
A similar situation applies to emission from an IsoBH accreting from clumps in the LIC (or cold neutral medium), although at the highest frequencies of SKA and with JWST, detection is possible within just tens of minutes of integration. Finally, emission from an IsoBH accreting from cold, dense clumps is expected to be easily detectable by all the telescopes considered, even with very short integration times.

Unfortunately, despite the promising observability, we have to rule out the presence of an IsoBH in the local cold dense clumps. 
Such clumps occupy only about 0.1\% of the volume within a $50\, \pc$ radius around the Sun. We estimated earlier that there should be only about 35 IsoBHs in this region. Therefore, the probability of an IsoBH being located inside a cold dense clump is just about 0.035. Moreover, if an IsoBH were indeed located in the local ISM so close to us, and had such a strong emission as predicted, it most likely would have already been identified.

\subsection{IsoBH at 200 pc from the Sun}\label{sec:sol200}

\begin{figure*}
\vspace{-0.0cm}
\centering
\begin{tabular}{cc}
\includegraphics[width=0.99\columnwidth]{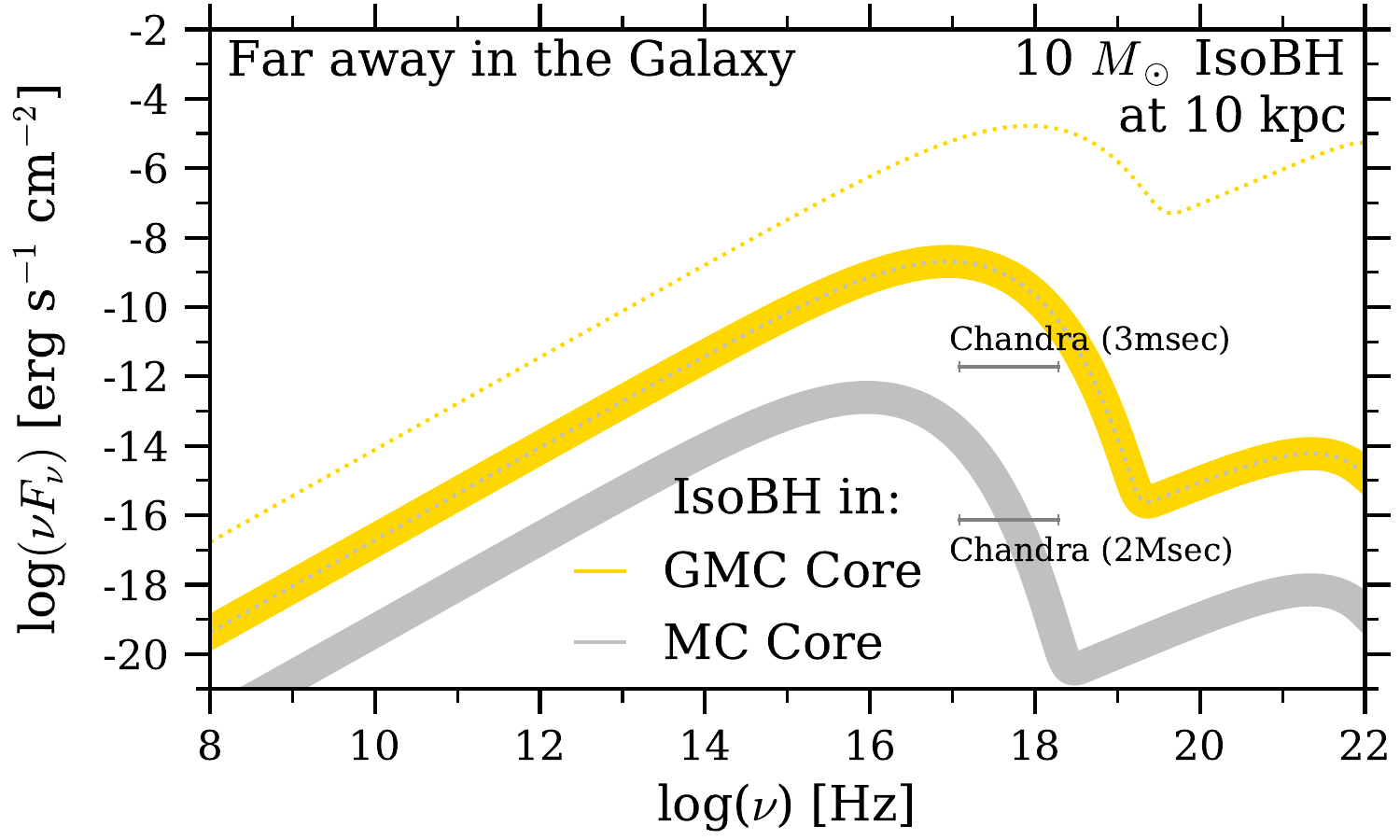} & 
\includegraphics[width=0.99\columnwidth]{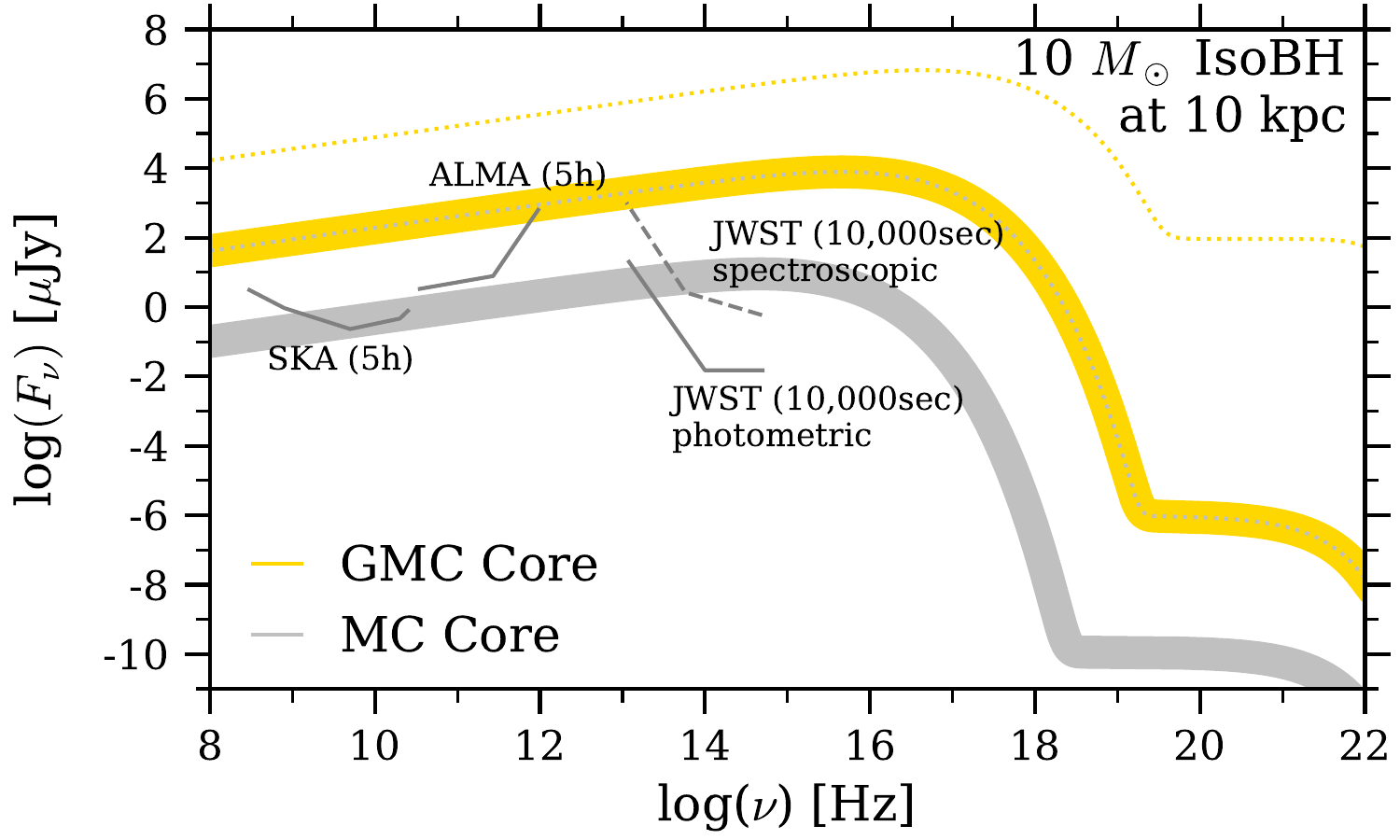}
\end{tabular}
\caption {
Estimated emission spectra of an IsoBH located 10 kpc from the Sun and accreting from the core of a GMC with densities $10^6 \ \cm^{-3}$ (yellow), and a core of an ordinary MC with density $10^4 \ \cm^{-3}$ (grey).
The thin dotted line shows the expected emission spectrum from an IsoBH accreting at the Bondi rate $(\lambda=1)$, representing the upper limit of the emission. The thick solid lines below the dotted
lines, shown in corresponding colors, represent the spectra from the IsoBH accreting at a more realistic 1\% of the Bondi rate $(\lambda=0.01)$. The exaggerated thickness illustrates large uncertainties in the accretion-model parameters. For convenience, we show the spectra in two different units ($\nu F_\nu$ and $F_\nu$) commonly used in different wavelength regions. 
Note that the dotted gray line overlaps with the thick yellow line. Thin dark gray lines indicate the observational sensitivities of Chandra, JWST, ALMA, and SKA, achievable within the integration times listed in brackets.
} 
\label{fig:cloudcore}
\end{figure*}

We expect about 1500 IsoBHs within 200 pc of the Sun. 
In this region, we have the following ISMs: coronal gas, warm partially ionized gas, cold neutral medium clumps within the warm medium, cold dense clouds, and MCs. 
There are several MCs within 200 pc \citep{zucker2021}: the Taurus MC located at $\sim 140 \ \pc$ (e.g. \citealt{2010ApJ...721..686P}), the Chamaeleon MC complex located at $\sim 180\ \pc$ \citep{Jordan2018,Galli2021}, 
the Corona Australis MC located at $\sim 130\ \pc$ \citep{Bracco2018}
and the Eos MC located at $\sim 94$ pc \citep{2025NatAs.tmp..108B}.

Figure \ref{fig:solar200} shows the estimated emission from 
our proto-typical IsoBH 
located at a distance of $200 \ \pc$ from the Sun and accreting from  the molecular cloud (grey), cold dense clouds (cyan), and cold neutral medium i.e. clumps within warm medium (orange).
IsoBHs accreting from the coronal gas and warm partially ionized medium were either undetectable or barely detectable already at 50 pc, so we omit discussing them here. 

Comparing the estimated spectra with the observational sensitivities of Chandra, SKA, ALMA, and JWST, we see that detecting the emission from an IsoBH located in cold neutral medium  can be achieved only with substantial investment of observational time, namely with several hours of integration with SKA and ALMA, or tens of minutes with JWST's highest frequencies.

The Taurus, Chamaeleon and Corona Australis MCs have average densities of $\sim 10^2-10^3\,  \cm^{-3}$, with core regions reaching densities of $\sim 10^4\, \cm^{-3}$ \citep{2010ApJ...721..686P, Hayashi2019, Pattle2025}. The cores occupy only about 1-2\% of the total volume of these MCs (e.g., \citealt{Pattle2017}). 
In contrast, the Eos MC has a much lower average density---between that of the Warm Medium and the Cold Neutral Medium---which makes it barely detectable, even with long integrations in JWST. For this reason, we do not discuss Eos cloud as a separate case.

Taking these densities into account, we find that the emission from an IsoBH accreting material from cold dense clouds---or from MCs such as the Taurus, Chamaeleon or Corona Australis--- is detectable with all telescopes with integration times from milliseconds to hours, depending on the observing frequency.

Cold dense clumps occupy around 0.1\% of the volume of the ISM within 200 pc from the Sun, we therefore expect about 1.5 IsoBHs located in such regions. MC occupy about 1\% of the ISM and therefore we expect about 15 IsoBHs located in this region. Taurus MC core occupy $\sim1-2$\% of its volume (e.g. \citealt{Pattle2017}), and therefore the probability of having an IsoBH within the core of Taurus MS is only $\sim 0.2,$ i.e. somewhat unlikely. Average density of the local MC is $\sim 10^2 - 10^3 \, \cm^{-3}$  \citep{2010ApJS..186..111L}, which is close to the density we used for cold dense medium. So we can conclude that out of 1500 IsoBHs within 200 pc from the Sun, about 16.5 are expected to be inside MCs or cold dense clouds and therefore can be detectable. Note that, when discussing detectability we do not consider extinction corrections. True detectability would depend on the exact location of the IsoBH, and accordingly, the calculations must account for the expected extinctions. We also note that the frequencies at or below submillimiter ($\sim 3 \times 10^{11} \, \mathrm{Hz}$) are less affected by dust extinction.
 
\subsection{In the rest of the galaxy}\label{sec:rest_gal}

At large distances from the Sun, IsoBHs are only detectable if they are accreting from the densest types of ISM---MCs and GMCs. The term GMC is generally used to describe molecular clouds that have a total mass exceeding $10^5 \Msun$ and a large spatial extent. Orion A and B molecular cloud complex---located at $\sim 400$ pc \citep{2008hsf1.book..459B}---is considered our closest GMC.

Figure \ref{fig:cloudcore} shows the estimated emission from $10 \Msun$ IsoBH moving through ISM with $50 \, \kms$ located at a distance of $10 \ \mathrm{kpc}$ from the Sun and accreting from a core of a GMC (yellow) and a core of an ordinary MC (grey). We used a typical density inside MC core to be $\sim 10^4 \ \cm^{-3}$ and inside GMC core to be $\sim 10^6 \ \cm^{-3}$ \citep{2010ApJS..186..111L}.

Comparing the estimated spectra with the observational sensitivities of Chandra, SKA, ALMA, and JWST, we see that the emission from an IsoBH accreting from the core of an ordinary MC (as well as from peripheral region of GMC) located at 10 kpc can be detected in each telescope with multiple hours of integration, while emission from an IsoBH accreting from cores of GMCs can be detected in each telescope with shorter integration times (except in high frequency bands of ALMA where several hours of integration would be required.)

MCs occupy about 1\% of our Galaxy by volume, implying that out of the $10^8$  IsoBHs in our Galaxy, about $10^6$ are located within MCs. 
The peripheral regions of MCs, away from their dense cores, have densities of $\sim 10^2-10^3 \ \cm^{-3}$ \citep{2015ARA&A..53..583H}.
As a result, IsoBHs in these peripheral regions are not detectable at large distances. Only IsoBHs that are situated in the dense cores of MCs---which typically occupy $\sim 0.1 - 1$\% of their volume \citep{2020SSRv..216...76B}---are detectable at large distances of $\sim 10$~kpc (see Figure~\ref{fig:cloudcore}), and in denser peripheral regions of the GMCs. 

Of course, the detectability is a function of proximity as well as the strength of the emission, so IsoBHs located sufficiently close to us can be detected even in the peripheral regions of MCs. 
As mentioned above, Orion---the closest GMC---is only 400 pc away, so the estimated emission from an IsoBH in Orion would be 625 times greater than that shown in Figure \ref{fig:cloudcore}. Consequently, detecting an IsoBH in Orion may require only short integration times for detection. Therefore, Orion provides a potentially promising target for detection of IsoBHs. 

\section{Identification}

\begin{figure*}
\vspace{-0.0cm}
\centering
\begin{tabular}{cc}
\includegraphics[width=0.99\columnwidth]{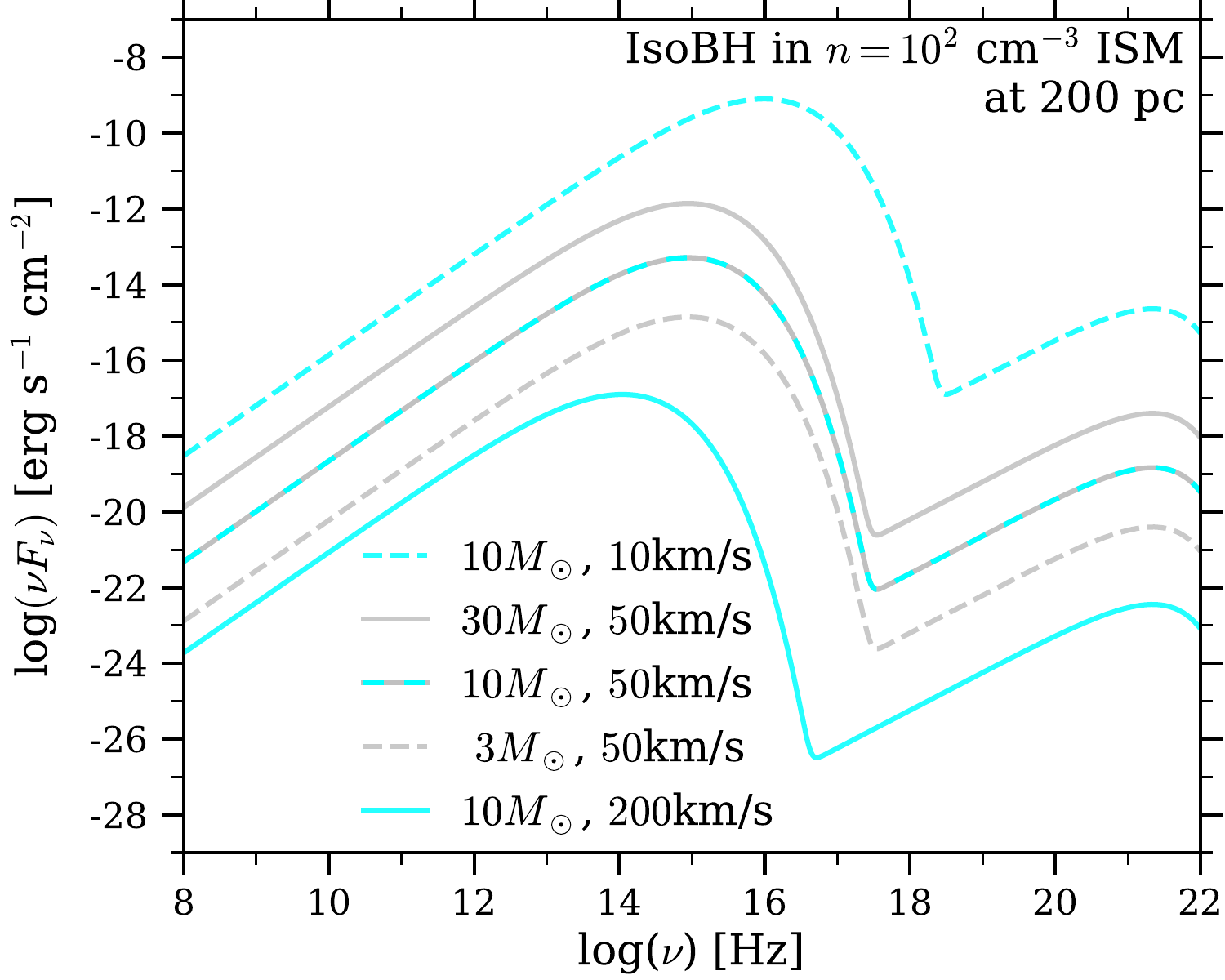} & 
\includegraphics[width=0.99\columnwidth]{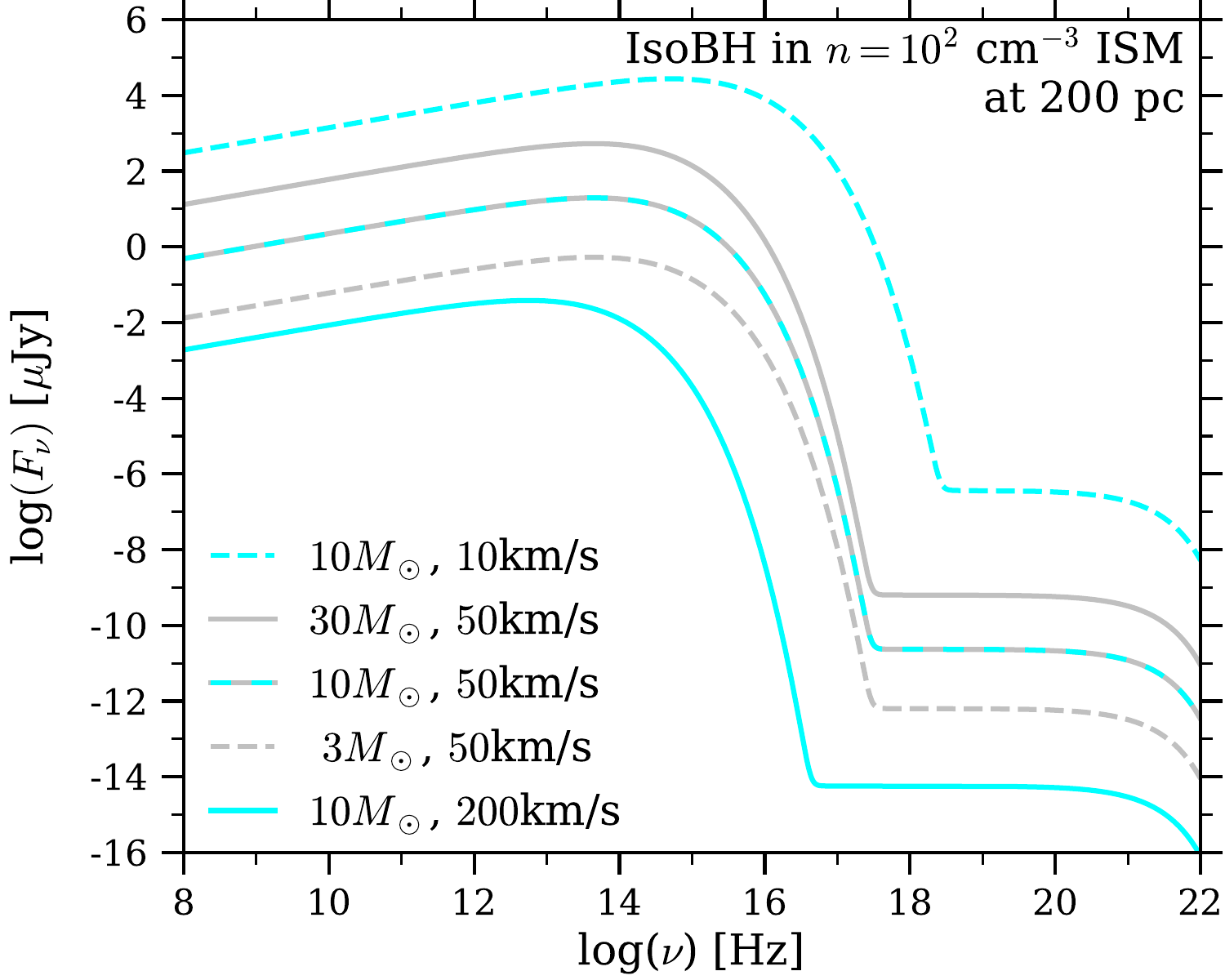}
\end{tabular}
\caption {
Estimated emission spectra of an IsoBH with different masses and velocities located 200 pc from the Sun, accreting from the ISM with a density of $100 \, \cm^{-3}.$ We show how the spectra change with IsoBH mass: $3 \, \Msun$ (dashed gray line), $10 \, \Msun$ (dashed gray-cyan line), and $30 \, \Msun$ (gray line) all moving at $50 \, \kms.$ We also show the effect of varying velocity for a $10 \, \Msun$: $10 \, \kms$ (dashed cyan line), $50 \, \kms$ (dashed gray-cyan line), and $200 \, \kms$ (cyan line). In each case, the solid line corresponds to the highest value of the parameter being varied. The IsoBH is assumed to accrete at 1\% of the Bondi rate $(\lambda=0.01)$. 
For convenience, the spectra are presented in two different units ($\nu F_\nu$ and $F_\nu$) commonly used in different wavelength regions. 
} 
\label{fig:mv}
\end{figure*}

In the previous sections, we showed that in many cases where IsoBHs accrete from various types of ISM, they are expected to be readily detectable with current telescope capabilities. However, merely detecting photons from an IsoBH does not ensure its identification as such. Fortunately, IsoBHs exhibit distinctive variability and spectral properties that, in principle, can be used to differentiate them from other types of astrophysical objects.

Accreting BHs are variable sources of electromagnetic emission. The characteristic timescale of their variability is comparable with the period of the innermost stable circular orbit. For an IsoBH with a mass of $10 \Msun$, this variability timescale $\tau_{ch} \simeq 7\, \mathrm{ms}$, and the variability is not expected to be periodic. 
We infer some additional details of this variability below, drawing on observations of the variability of Sagittarius A$^*$
(e.g., \citealt{2021ApJ...917...73W,2006ApJ...641..302M,2022agn..book...65M}), rescaled to the physical parameters relevant for IsoBHs. At the peak of its spectrum, the IsoBH emission is expected to vary smoothly over timescales $\sim \tau_{ch}$, deviating from its average flux by a factor of a few. 
This emission does not exhibit true flaring behavior, characterized by sharp contrasts between base and peak levels; rather, it fluctuates gradually—--so to speak, it ``wanders."
Due to relatively strong variability on timescales of $\tau_{ch}$, it is difficult to accurately characterize longer-term variability on timescales of $\sim 10^3 \tau_{ch}$ or more, though such variability is expected to be present. 
At lower energies, the emission becomes less variable on shorter timescales of $\sim 10 \tau_{ch}$,
while remaining variable on longer timescales of $\gtrsim 10^3 \tau_{ch}$.
Toward higher energies, the emission becomes more variable, exhibiting pronounced flaring activity, with flares that are taller and more distinct relative to the baseline emission.

The spectra shown in Figures \ref{fig:solar50}-\ref{fig:cloudcore} represent average emission profiles. As discussed above, the variability of IsoBHs is not expected to be periodic, so folding the data as done for pulsar searches would not be applicable here. Therefore, to detect the variability, one must directly observe the IsoBH with high time resolution: at millisecond cadence near the peak of the BH emission, or on timescales ranging from $0.1$ to few seconds at frequencies one to two decades away from the peak energies. 
However, such observations are only feasible for the brightest IsoBHs—--those located in the cores GMCs throughout the Galaxy (Figure \ref{fig:cloudcore}), or in the densest regions of the local ISM, such as the local cold dense clouds, 
the cores of Orion, Taurus, Perseus, Chamaeleon and Corona Australis MCs or other dense regions within MCs.  It is important to note that integrating the emission over timescales of several hours would effectively average out the variability, rendering it undetectable.

We expect hundreds to $\sim 1000$ IsoBHs to be present in the cores of GMCs (Section \ref{sec:rest_gal}). Unfortunately, these regions are compact, have complex structures, are located at large distances, heavily extinct, and somewhat busy with other sources of emission unrelated to IsoBHs, e.g., star formation. Identifying an object with weak emission, whose only distinguishing feature is chaotic millisecond variability in these crowded regions, would be extremely difficult and somewhat unlikely.

We expect $\sim 16$ IsoBHs detectable with millisecond cadence in the Solar neighborhood within 200pc (Section \ref{sec:sol200}).
They would be located in the 
MCs like Taurus, Chamaeleon and Corona Australis
and in the cold dense clouds 
scattered in the ISM\footnote{We know the multiple cold dense clouds in the immediate solar surrounding, beyond which they are hard to detect \citep{2008ApJ...673..283R}. However we do not presume the ISM surrounding the Sun to be very unique, and therefore expect such cold dense clouds to be scattered in the ISM all over the Galaxy.}. 
Although we expect only a relatively small number of such IsoBHs, their proximity and well-constrained locations---such as the cores of the MCs---make them strong candidates for detecting variability on millisecond to second timescales.

The majority of IsoBHs, as evident from the spectra presented in Figures \ref{fig:solar50}-\ref{fig:cloudcore}, can be detected only with minutes to hours-long integrations. No variability can be detected in IsoBHs which require long integrations for their detection. This is the case for IsoBHs located in the galactic cold neutral medium of the local ISM, peripheral regions of MCs, and other dense ISM. Identifying such IsoBHs can be done through spectral properties, which for robustness requires multi-telescope observations. This is possible in case-by-case studies of individual objects of interest but might be impractical for designated searches due to large telescope time commitment. In each individual telescope, IsoBH emission would appear like an object with a relatively featureless spectrum, except affected by local ISM chemistry, and those will be unidentified or sometimes misidentified in each individual wavelength survey. It is more likely to identify IsoBHs in the Solar neighborhood due to its proximity than far away in the Galaxy.

So far, we have discussed the observability of a prototypical IsoBH with mass and velocity relative to the ISM given in Equation \ref{eq:mv}. These values are close to the peak of the 
Galactic BH mass distribution and fall within the mid-range of the expected, though uncertain, velocity distribution. Figure \ref{fig:mv} illustrates the dependence of the expected emission spectra for masses ranging from a few $\Msun$ to $30 \, \Msun$, and velocities relative to the surrounding ISM ranging from $ 10 \, \kms$ to $ 200 \, \kms$ (see discussion in the Introduction).
We see that the dependence on mass spans about $\pm 1.5$ orders of magnitude, with more massive IsoBHs being easier to detect than less massive ones. 
This range of uncertainty across the full expected IsoBH mass spectrum is comparable to that expected from the combined uncertainties in the accretion parameter $\lambda$
in Equation \ref{eq:lambda}  and variations in the surrounding ISM conditions. The dependence on velocity is very strong, with slower-moving black holes more readily detectable than faster ones. This is not surprising, as Equation \ref{eq:Bnum}  shows that the emission scales inversely as the cube of the velocity.
Effectively, a change in the emission spectrum due to a change in velocity by a factor of $\xi_v$ is equivalent to a change in the surrounding density by a factor of $\xi_v^{-3}.$ In other words, a  decrease in IsoBH velocity by a factor of just $2.15$ results in an increase in the expected emission
equivalent to the increase in the surrounding density by a factor of 10 -- potentially making detection possible even in a low-density ISM. The reverse is also true: higher velocities suppress detectability.

The velocity distribution of the IsoBHs is highly uncertain \citep{mandel2016,repetto2012}. If the distribution is skewed toward lower velocities, the detectability and probability of identifying IsoBHs would 
increase by many folds. 
Conversely, if the distribution is skewed toward higher velocities, detection would become considerably more challenging.  

\section{Detection Prospects and Conclusions}

We find that photons emitted by IsoBHs accreting from various types of ISM  routinely reach us and are readily detectable with current telescopes such as SKA, ALMA, JWST, and Chandra (Figures \ref{fig:solar50}-\ref{fig:cloudcore}). However, the  
scarcity of confirmed IsoBHs is primarily due to the 
challenges involved in reliably identifying them as accreting BHs. IsoBHs located in particularly dense ISM, such as the core of GMCs, are expected to be sufficiently luminous to be detectable across the Galaxy on millisecond timescale which can aid their identification as stellar mass BHs. The estimated number of such IsoBHs ranges from a few hundred to $\sim 1000$. However, the dense cores of GMCs are complex, crowded, distant, and heavily obscured regions, making it difficult to isolate relatively faint, chaotically variable sources on millisecond timescales from the background.
Conversely, less crowded and less complex regions of the ISM typically have much lower densities, resulting in significantly weaker IsoBH emission and consequently harder to detect.

The majority of IsoBHs can be detected with integration times on the order of hours, making the detection of their characteristic variability unlikely. This applies to IsoBHs located in the Galactic cold neutral medium of the local ISM, peripheral regions of MCs, and other moderately dense ISM environments. Identifying such IsoBHs would therefore rely on their spectral properties, which necessitates multi-telescope observations. In data from individual telescopes, IsoBH emission is expected to manifest as a relatively featureless spectrum, increasing the risk of these sources remaining unidentified or being misclassified.

The most promising regions to search for IsoBHs through their electromagnetic emission appear to be within the local ISM, a few hundred parsecs from the Sun, including the outskirts of the nearby MCs. 
Their relative proximity allows IsoBHs to be detectable even in lower-density environments and makes them more easily identifiable compared to similar sources located in the denser, more distant regions of the Galaxy.

The strength of the IsoBH emission depends strongly on their velocity relative to the ISM (Figure \ref{fig:mv}). While the expected velocity distribution of IsoBHs remains uncertain, observations of these objects could provide valuable constraints. This is because detectability dramatically increases as $v_\bullet$ decreases, and decreases as $v_\bullet$ increases.

Upcoming Vera C. Rubin Observatory's Legacy Survey of Space and Time (LSST) offers significant promise for detection and identification of IsoBHs in the near future. LSST will monitor the entire observable sky every few days, reaching sensitivity down to magnitude 25 per visit, corresponding to approximately $0.36 \, \mu \mathrm{Jy}$. Each visit will consist of two 15 sec integrations, achieving a photometric precision of 10 
millimagnitudes per visit
(\citealt{Bianco2022}, and \url{https://rubinobservatory.org/for-scientists/rubin-101/key-numbers})
The sensitivity of LSST is sufficient to detect emission from IsoBHs located in dense regions of the ISM (cold sense clumps and MCs) within a few hundred parsecs, even in a single visit. Although the integration time is relatively large, it remains compatible with the expected timescales of long-term IsoBH variability of $\sim 1000 \, \tau_{ch}$, allowing for the potential detection of flux changes on the order of several tens of percent.

It is most likely that numerous IsoBHs are already present in existing catalogs but have not been recognized as such. 
In this work, we have provided guidelines designed to support further detections of IsoBHs and their accurate identification using current and upcoming facilities.

\section*{Acknowledgements}
We are grateful to Dave Meyer, Shigeo Kimura, Daniel Palumbo, Sasha Philippov, Roman Rafikov, Roger Blandford, Jason Wang, Ramesh Narayan, Hannah Bish, and the anonymous referee for their valuable suggestions.

\bibliography{Bibliography}{}
\bibliographystyle{aasjournal}
 
\end{document}